\documentclass[twocolumn,showpacs,preprintnumbers,amsmath,amssymb,floatfix]{revtex4}
\usepackage{amsmath,amsfonts,amssymb,amsthm,epsfig,epstopdf,url,array}
\usepackage{graphicx}
\usepackage{dcolumn}
\usepackage{bm}

\newcommand{\beq}{\begin{equation}}
\newcommand{\eeq}{\end{equation}}
\newcommand{\bey}{\begin{eqnarray}}
\newcommand{\eey}{\end{eqnarray}}

\begin{document}

\title{ Isotropic star in low-mass X-ray binaries and X-ray pulsars}

\author{ Mehedi Kalam}
\email{kalam@iucaa.ernet.in} \affiliation{Department of
Physics, Aliah University, Sector - V , Salt Lake,  Kolkata -
700091, India}

\author{Sk. Monowar Hossein}
\email{hossein@iucaa.ernet.in} \affiliation{Department of
Mathematics, Aliah University, Sector - V , Salt Lake,  Kolkata -
700091, India}

\author{Sajahan Molla}
\email{sajahan.phy@gmail.com} \affiliation{Department of Physics,
Aliah University, Sector - V , Salt Lake,  Kolkata, India}

\date{\today}

\begin{abstract}
We present a model for compact stars in the low mass X-ray binaries(LMXBs) and X-ray pulsars using a metric given by John J. Matese and Patrick G. Whitman \citep{Matese and Whitman1980}. Here the field equations are reduced to a system of two algebraic equations considering the isotropic pressure. Compact star candidates 4U 1820-30(radius=10km) in LMXBs, and Her X-1(radius=7.7km), SAX J 1808.4-3658(SS1)(radius=7.07km) and SAX J 1808.4-3658(SS2)(radius=6.35km) in X-ray pulsars satisfy all the energy conditions, TOV-equation and stability condition. From our model, we have derived mass($M$), central density($\rho_{0}$), suface density($\rho_{b}$), central pressure($p_{0}$), surface pressure($p_{b}$) and surface red-shift($Z_{s}$) of the above mentioned stars, which are very much consistant with the observed/reported datas\citep{N. K. Glendenning1997,Gondek2000}. We have also observe the adiabatic index($\gamma>\frac{4}{3}$) of the above steller objects.
\end{abstract}

\maketitle

\section{Introduction}

Compact stars in the low-mass X-ray binaries(LMXBs) and X-ray pulsars are of great attention for the last few years. In the LMXBs and X-ray pulsars, possible compact objects are neutron stars or strange stars. Neutron stars are composed of neutrons, while strange stars are of quark or strange matter. Chandra X-ray Observatory data has revealed LMXBs and X-ray pulsars in many distant galaxies. Few LMXBs and X-ray pulsars have been detected in our Milkyway galaxy. A low-mass X-ray binary or X-ray pulsar emits radiation mostly in X-rays. LMXBs have the orbital period range from ten minutes to several hundred days, while that of X-ray pulsars ranges from as little as a fraction of a second to as much as several minutes. 4U 1820-30(radius=10km) is a possible neutron star or strange star in LMXBs and SAX J 1808.4-3658(SS2)(radius=6.35km), SAX J 1808.4-3658(SS1)(radius=7.07km) and Her X-1(radius=7.7km) are in X-ray pulsars.\\
The first solution of Einstein's field equation was given by Schwarzschild\citep{Schwarzschild1916} to describe the interior structure of a compact object.Tolman\citep{Tolman1939} also proposed a set of static isotropic (a set of eight) solutions for a sphere of fluid and these solutions still most important exact interior solution of the gravitational field equations. John J. Matese and Patrick G. Whitman \citep{Matese and Whitman1980} gave three (M-W I, M-W II and  M-W III) solutions for ultra-dense objects. According to Delgaty and Lake \citep{Lake1998} only M-W I solution satisfy all the required conditions. Motivated by these results, we are specially interested for analytical modeling of the compact stars 4U 1820-30 in LMXBs and SAX J 1808.4-3658(SS2), SAX J 1808.4-3658(SS1), Her X-1 in X-ray pulsars. Several researcher investigated compact stars analytically or numerically\citep{Rahaman2012a,Kalam2012a,Hossein2012,Rahaman2012b,Kalam2013b,Kalam2013c,Lobo2006,Egeland2007,Dymnikova2002,T. Harko2013,D.E. Barraco and V.H. Hamity2002,Richard C. Tolman1939,H. Knutsen1988,S. Chandrasekhar1964,J. M. Bardeen1966}. Here we compare our mesurements of mass, radius, central density, surface density, equation of state(EOS), central pressure, surface pressure, surface red-shift, adiabatic index and compactness with the strange stars 4U 1820-30 in LMXBs and SAX J 1808.4-3658(SS2), SAX J 1808.4-3658(SS1), Her X-1 in X-ray pulsars, and it is found to be consistant with the available datas\citep{N. K. Glendenning1997,Gondek2000,Hossein2012,Kalam2013b}. We also check the stability (sound velocity $\leq1$ ) of our model.\\
 We organise the paper as follows:\\
In Sec II we provide the basic equations in connection to the compact star using M-W I metric. In Sec. III we discuss the physical behavior of the star, namely-
(A) Density and pressure of the star, (B) Matching conditions, (C) TOV equations, (D) Energy
conditions, (E) Stability analysis, (F) Adiabetic index, (G) Mass-radius relation and surface red-shift are discussed in different Sub-sections. The article concludes with derived results and a brief discussion.

\section{Interior solution}
Let the interior space-time of a compact star be described by a spherically symmetric metric of the form \pagebreak

\begin{eqnarray}
ds^{2} = -\left(1+a\frac{r^{2}}{R^{2}}\right) \left(A \sin z +B\cos z\right)^{2} dt^{2}  \nonumber\\
+\left(1+2a\frac{r^{2}}{R^{2}}\right)dr^{2} +r^2d\Omega^{2}. \label{eq1}\\
where,  z = \sqrt {1+2a\frac{r^{2}}{R^{2}}}-\tan^{-1}\sqrt {1+2a\frac{r^{2}}{R^{2}}}. \nonumber
\end{eqnarray}
and $a$,$A$,$B$,$R$ are constants. Such type of metric (1) was proposed by John J. Matese and Patrick G. Whitman\citep{Matese and Whitman1980}. 
We also assume the energy-momentum tensor as
\begin{equation}
               T_\nu^\mu=  ( -\rho ,p, p, p)
         \label{Eq2}
          \end{equation}

where $\rho$ is the energy-density and $p$ is the isotropic pressure.

Solutions of the Einstein's field equations for the metric (1) accordingly are obtained as (considering $c=1,G=1$)
\begin{widetext}
\begin{eqnarray}
8\pi  \rho &=& \left(1+2a\frac{r^{2}}{R^{2}}\right)^{-1} \left[\frac{4a}{\left(R^{2}+2ar^{2}\right)}-\frac{1}{r^{2}}\right] + \frac{1}{r^2}\label{eq2}\\
8\pi  p &=& \left(1+2a\frac{r^{2}}{R^{2}}\right)^{-1}
\left[\frac{1}{r^2} +\frac{2a}{\left(R^{2}+ar^{2}\right)}+
\frac{2a\sqrt {1+2a\frac{r^{2}}{R^{2}}}}{\left(R^{2}+ar^{2}\right)}\frac{\left(A \cos z -B\sin z\right)}{\left(A \sin z +B\cos z\right)}\right]
-\frac{1}{r^2}.\label{eq3}
\end{eqnarray}
\end{widetext}

\section{Exploration of Physical properties}
In this section we will investigate the following features of the compact stars:

\subsection{Density and Pressure Behavior of the star}


To check whether at the centre, the matter density and pressure dominates or not,
we see the following:
\begin{equation}
\frac{d\rho}{dr}~< ~0,\nonumber\\
~
\left(\frac{d\rho}{dr}\right)_{r=0}~=0, \nonumber\\ 
~\left(\frac{d^2 \rho}{dr^2}\right)_{r=0} < 0\nonumber
\end{equation}
and 
\begin{equation}
\frac{dp}{dr}  < 0,\nonumber\\
~\left(\frac{dp}{dr}\right)_{r=0}~=0, \nonumber\\ 
\left(\frac{d^{2}p}{dr^{2}}\right)_{r=0}  < 0.\nonumber
\end{equation}
Clearly, at the centre, the density and pressure of the star is maximum and it decreases
radially outward. \\
Thus, the energy density and the pressure are well behaved in the interior of the stellar
structure. 
Variations of the energy-density and  pressure in our propose model
have been shown  for different stars in FIG.~1 and FIG.~2, respectively. Here also pressure and energy-density gradients is negative(see FIG.~3 and FIG.~4). Equation of state of the different strange stars at the stellar interior is shown in FIG.5. 

\begin{figure}[htbp]
\centering
\includegraphics[scale=.4]{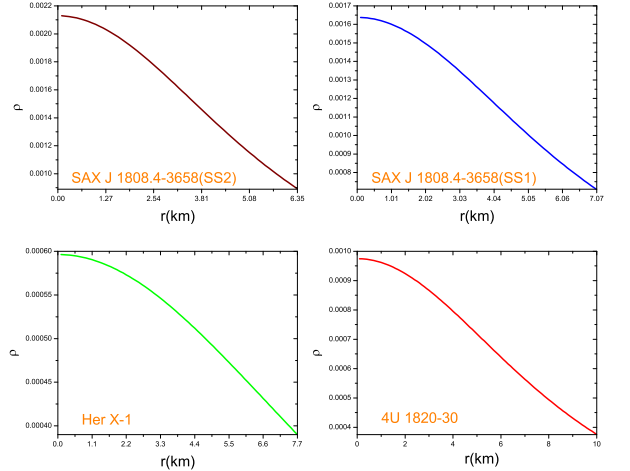}
\caption{Variation of the energy-density($\rho$) at the stellar interior of SAX J 1808.4-3658(SS2)(radius=6.35km), SAX J 1808.4-3658(SS1)(radius=7.07km), Her X-1(radius=7.7km), and 4U 1820-30(radius=10km).}
\label{fig:1}
\end{figure}

\begin{figure}[htbp]
\centering
\includegraphics[scale=.4]{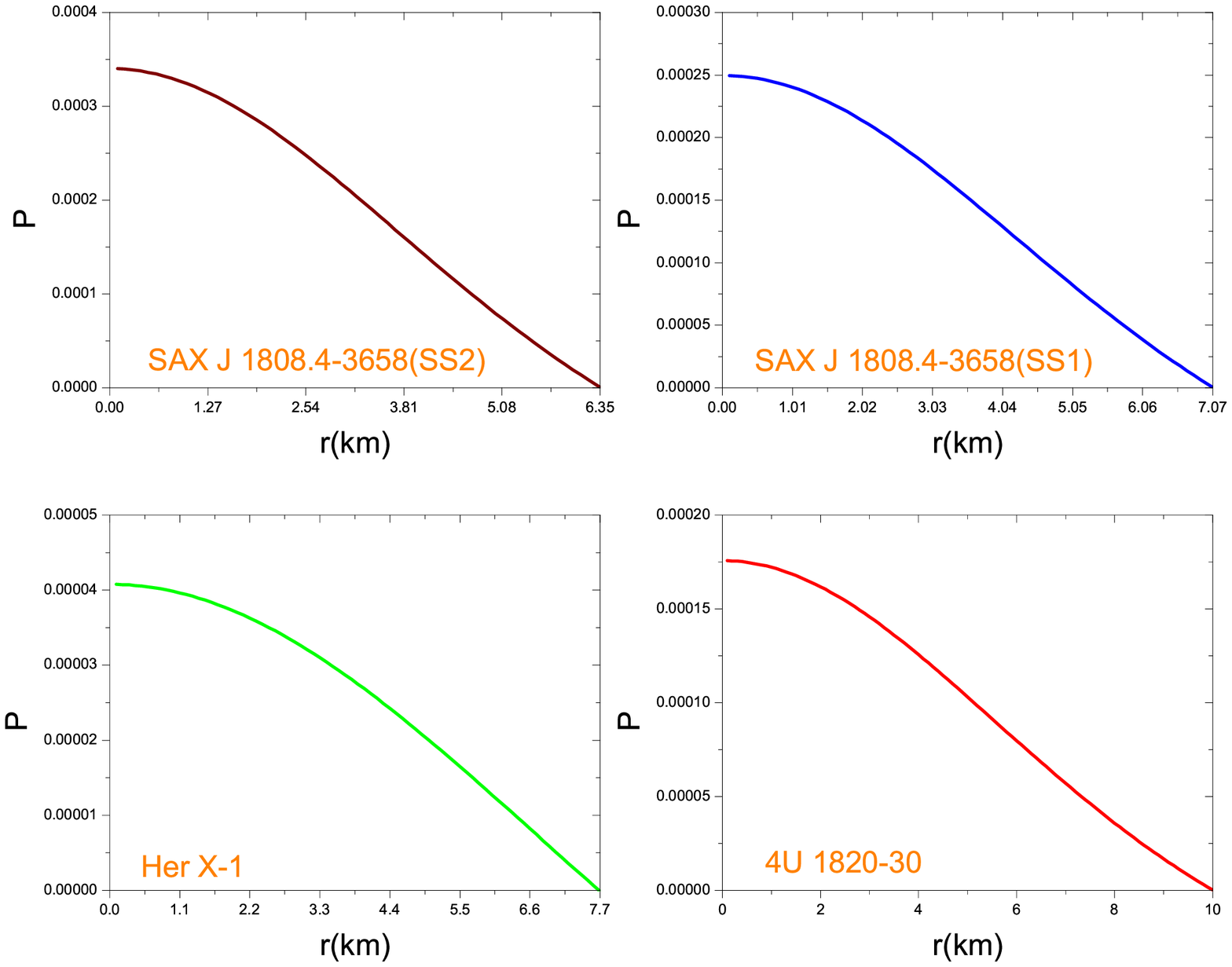}
\caption{Variation of the pressure ($p$) at the stellar interior of SAX J 1808.4-3658(SS2)(radius=6.35km), SAX J 1808.4-3658(SS1)(radius=7.07km), Her X-1(radius=7.7km), and 4U 1820-30(radius=10km).}
\label{fig:2}
\end{figure}

 \begin{figure}[htbp]
\centering
\includegraphics[scale=.4]{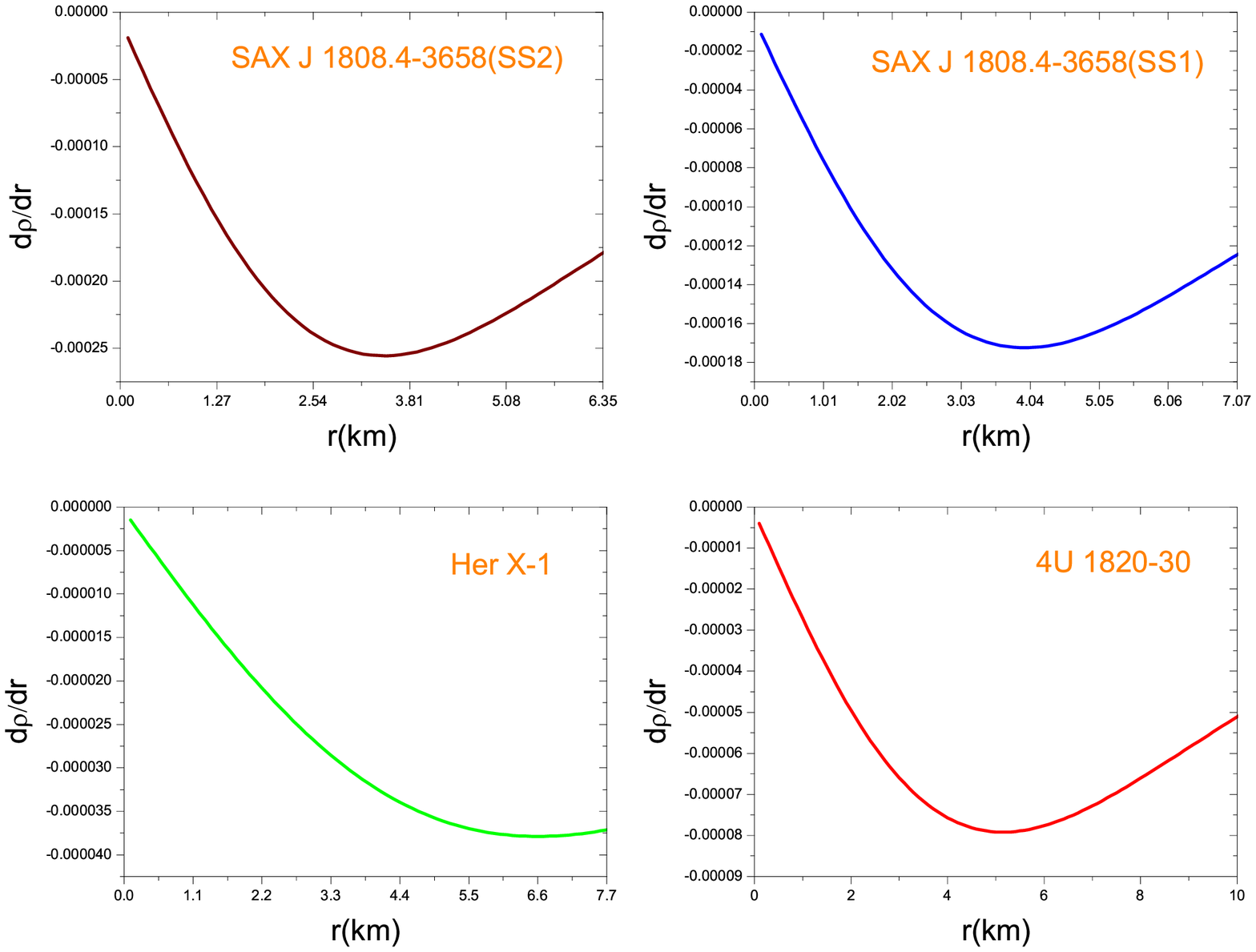}
\caption{Variation of the energy-density gradients($\frac{d\rho}{dr}$) at the stellar interior of SAX J 1808.4-3658(SS2)(radius=6.35km), SAX J 1808.4-3658(SS1)(radius=7.07km), Her X-1(radius=7.7km), and 4U 1820-30(radius=10km).}
\label{fig:3}
\end{figure}

\begin{figure}[htbp]
\centering
\includegraphics[scale=.4]{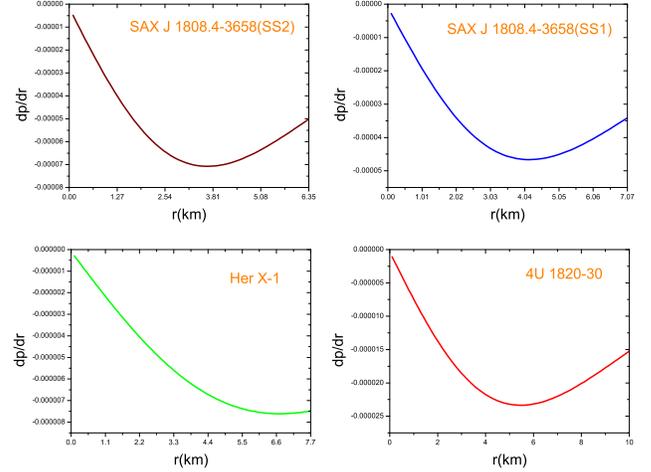}
\caption{Variation of the pressure gradients($\frac{dp}{dr}$) at the stellar interior of SAX J 1808.4-3658(SS2)(radius=6.35km), SAX J 1808.4-3658(SS1)(radius=7.07km), Her X-1(radius=7.7km), and 4U 1820-30(radius=10km).}
\label{fig:4}
\end{figure}

\begin{figure}[htbp]
\centering
\includegraphics[scale=.4]{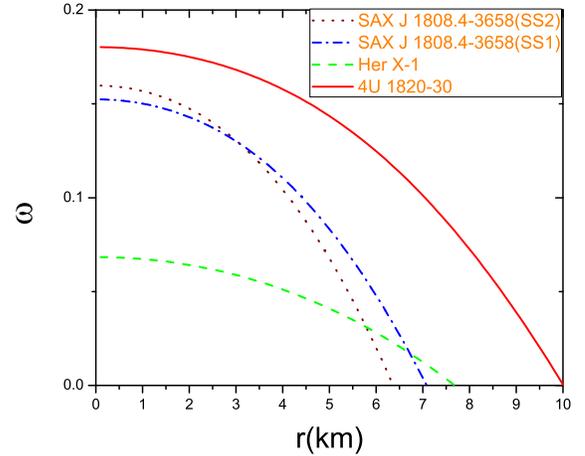}
\caption{EOS parameter($\omega$) of the strange stars at the stellar interior of SAX J 1808.4-3658(SS2)(radius=6.35km), SAX J 1808.4-3658(SS1)(radius=7.07km), Her X-1(radius=7.7km), and 4U 1820-30(radius=10km).}
\label{fig:5}
\end{figure}

\subsection{Matching Conditions}
Interior metric of the star should match to the Schwarzschild exterior metric (at the boundary $r=b$, where b is the radius of the star).
\begin{eqnarray}
ds^2 = - \left(1-\frac{2M}{r}\right)dt^2 +  \left(1-\frac{2M}{r}\right)^{-1}dr^2  \nonumber\\
+r^2(d\theta^2 +sin^2\theta d\phi^2). \label{eq1}
\end{eqnarray}

Assuming the continuity of the metric functions $g_{tt}$,~ $g_{rr} $ and $\frac{\partial g_{tt}}{\partial r}$ at the boundary, we get

\begin{eqnarray}
\left(1+a\frac{b^{2}}{R^{2}}\right) (A \sin z +B\cos z)^{2}  &=& \left(1 - \frac{2M}{b}\right),\label{eq13}
\end{eqnarray}

\begin{equation}
\left(1+2a\frac{b^{2}}{R^{2}}\right) = \left(1-\frac{2M}{b}\right)^{-1}  .\label{eq14}
\end{equation}

Now from equation ~(\ref{eq14}) , we get the compactification factor as
\begin{equation}
u~=~\frac{M}{b} = \frac{ab^{2}}{R^{2}+2ab^{2}}.\label{eq15}
\end{equation}

\subsection{TOV equation}
For isotropic fluid distribution, the generalized TOV equation is of the form
\begin{equation}
\frac{dp}{dr} +\frac{1}{2} \nu^\prime\left(\rho+ p\right)= 0.\label{eq18}
\end{equation}
Following \citet{Leon1993}, we rewrite the above TOV equation as
\begin{equation}
-\frac{M_G\left(\rho+p\right)}{r^2}e^{\frac{\lambda-\nu}{2}}-\frac{dp
}{dr} = 0, \label{eq19}
\end{equation}
where $M_G(r)$ is the gravitational mass inside a star of radius $r$ and is given by
\begin{equation}
M_G(r) = \frac{1}{2}r^2e^{\frac{\nu-\lambda}{2}}\nu^{\prime},\label{eq20}
\end{equation}
and $e^{\lambda(r)} = \left(1+2a\frac{r^{2}}{R^{2}}\right),$

which can easily be derived from the Tolman-Whittaker formula and Einstein's field equations. The modified  TOV equation describes the equilibrium condition for the compact star subject to the equlibrium condition  of effective gravitational($F_g$) and effective hydrostatic($F_h$) forces as:
\begin{equation}
F_g+ F_h  = 0,\label{eq21}
\end{equation}
where the force components are given by
\begin{eqnarray}
F_g &=& -\frac{1}{2}\nu^{\prime}\left(\rho+p\right)\label{eq22}\\
F_h &=& -\frac{dp}{dr}. \label{eq23}
\end{eqnarray}
FIG.~6 indicates the existence of static equilibrium configurations of the accounted stars.

\begin{figure}[htbp]
    \centering
        \includegraphics[scale=.4]{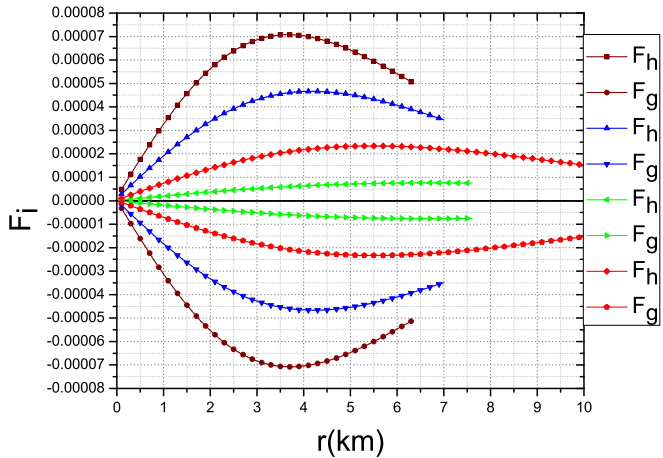}
        \caption{The Behaviors of gravitational($F_{g}$) and hydrostatic($F_{h}$) forces at the
stellar interior of the SAX J 1808.4-3658(SS2)(radius=6.35km), SAX J 1808.4-3658(SS1)(radius=7.07km), Her X-1(radius=7.7km),and 4U 1820-30(radius=10km).}
    \label{fig:6}
\end{figure}

\subsection{Energy conditions}
In our model all the energy conditations, namely, null energy condition(NEC), weak energy condition(WEC), strong energy condition(SEC) and dominant energy condition(DEC), are satisfied at the centre ($r=0$). We have evaluated the numerical values of the parameters $ A, a, R, B $ (see Table I) for the compact stars 4U 1820-30(radius=10km), Her X-1(radius=7.7km), SAX J 1808.4-3658(SS1)(radius=7.07km), SAX J 1808.4-3658(SS2)(radius=6.35km). We get the following satisfied energy conditions as:\\
(i) NEC: $p_{0}+\rho_{0}\geq0$ ,\\
(ii) WEC: $p_{0}+\rho_{0}\geq0$  , $~~\rho_{0}\geq0$  ,\\
(iii) SEC: $p_{0}+\rho_{0}\geq0$  ,$~~~~3p_{0}+\rho_{0}\geq0$ ,\\
(iv) DEC: $\rho_{0} > |p_{0}| $.\\
The value of $p_{0}$ and $\rho_{0}$ are also given in Table I.

\begin{table*}
\centering
\begin{minipage}{140mm}
\caption{Values of the parameters evaluated for
different compact stars.}\label{tbl}
\begin{tabular}{|lllllllll|}
\hline
Compact Star & a & R & A & B & $\rho_{0}$(km$^{-2}$) & $\rho_{b}$(km$^{-2}$) & $p_{0}$ (km$^{-2}$) & $p_{b}$ (km$^{-2}$) \\ \hline
SAX J 1808.4-3658(SS2) & 0.26 & 5.4 & 0.545 & 0.7 & 0.0021286 & 0.000892952 & 0.000340069 &  8.52288$\times10^{-7}$ \\
SAX J 1808.4-3658(SS1) & 0.2 & 5.4 & 0.3 & 0.4 & 0.0016374 & 0.000707956 & 0.000249588 & 4.97971$\times10^{-7}$ \\
Her X-1 & 0.2 & 8.95 & 0.186 & 0.42 & 0.0005961 & 0.000389864 & 0.000040751 & 2.49734$\times10^{-7}$ \\
4U 1820-30 & 0.2 & 7 & 0.43 & 0.5 & 0.0009744 & 0.000375735 & 0.000175619 & 1.71924$\times10^{-7}$ \\
\hline
\end{tabular}
\end{minipage}
\end{table*}

\subsection{Stability}
For a physically acceptable model, one expects that the velocity of
sound should be within the range  $0 \leq  v_s^2=(\frac{dp}{d\rho})
\leq 1$\citep{Herrera1992,Abreu2007}. \\

We plot the sound speeds for various stars in FIG.~7 and observe
that these parameter satisfy the inequalities condition $0\leq v_{s}^{2}\leq 1$ everywhere within the stellar objects.\\

These shows that our isotropic compact star model is stable.

\begin{figure}[htbp]

  \centering
        \includegraphics[scale=.4]{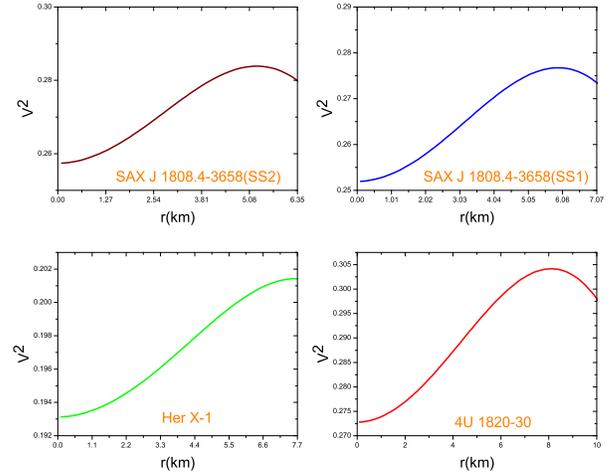}
       \caption{ Variation of the sound speed at the stellar interior SAX J 1808.4-3658(SS2)(radius=6.35km), SAX J 1808.4-3658(SS1)(radius=7.07km), Her X-1(radius=7.7km) and 4U 1820-30(radius=10km).}
   \label{fig:7}
\end{figure}

\subsection{Adiabatic Index}
The dynamical stability of the stellar model against the infinitesimal radial adiabatic perturbation was introduced by S. Chandrasekhar\citep{S. Chandrasekhar1964}. Later this stability condition was developed and applied to astrophysical cases by  J. M. Bardeen, K. S. Thorne, and D. W. Meltzer\citep{J. M. Bardeen1966}, H. Knutsen\citep{H. Knutsen1988}, M. K. Mak and T. Harko\citep{T. Harko2013}, gradually. Since the adiabatic index should be $\gamma = \frac{\rho+p}{p} \frac{dp}{d\rho}> \frac{4}{3}$  within the isotropic stable star, we plot the adiabatic index for our compact stars in FIG.~8 and observe that these parameter satisfy the condition $\gamma > \frac{4}{3}$ everywhere within the star.\\
\begin{figure}[htbp]

  \centering
        \includegraphics[scale=.4]{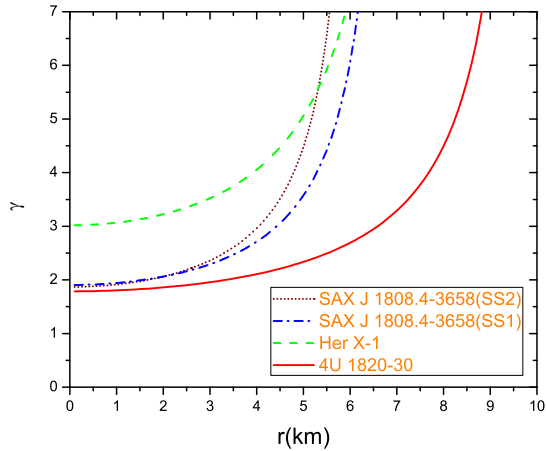}
       \caption{ Variation of the adiabatic index $\gamma $ at the stellar interior SAX J 1808.4-3658(SS2)(radius=6.35km), SAX J 1808.4-3658(SS1)(radius=7.07km), Her X-1(radius=7.7km) and 4U 1820-30(radius=10km).}
   \label{fig:8}
\end{figure}

\subsection{Mass-Radius relation and Surface Red-shift}
In this section, we study the maximum allowable mass-radius ratio. According to Buchdahl \citep{Buchdahl1959}, 
allowable mass-radius ratio should be $\frac{2M}{R} <
\frac{8}{9}$ for a static spherically symmetric perfect fluid sphere. Generalized expression for the mass in terms of the energy density $\rho$ can be expressed as
\begin{equation}
\label{eq34}
 M=4\pi\int^{b}_{0} \rho~~ r^2 dr = \frac{ab^{3}}{R^{2}+2ab^{2}}
\end{equation}

\begin{figure}[htbp]
\centering
\includegraphics[scale=.4]{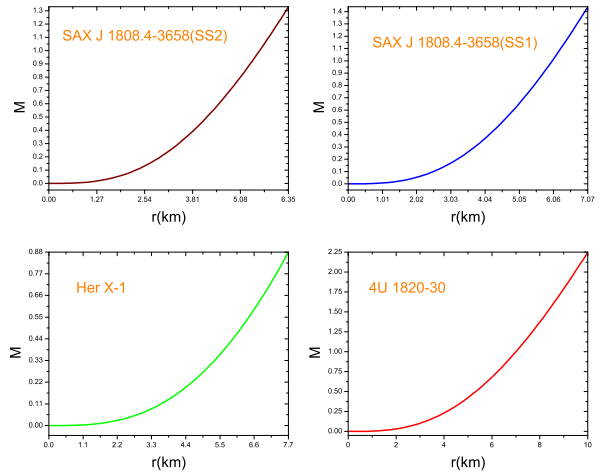}
\caption{Variation of M against radial parameter r of the star SAX J 1808.4-3658(SS2)(radius=6.35km), SAX J 1808.4-3658(SS1)(radius=7.07km), Her X-1(radius=7.7km) and 4U 1820-30(radius=10km).}
\label{fig:9}
\end{figure}
We note that a constraint on the maximum allowed mass-radius ratio in our case (FIG.~9) is
similar to the isotropic fluid sphere, i.e., $\frac{mass}{radius}
< \frac{4}{9}$ as obtained earlier\citep{Buchdahl1959}.\\ The compactness of the star is
given by
\begin{equation}
\label{eq35} u= \frac{ M(b)} {b}= \frac{ab^{2}}{R^{2}+2ab^{2}}
\end{equation}
FIG.~10 shows the variation of compactness of the stars.

\begin{figure}[htbp]
\centering
\includegraphics[scale=.4]{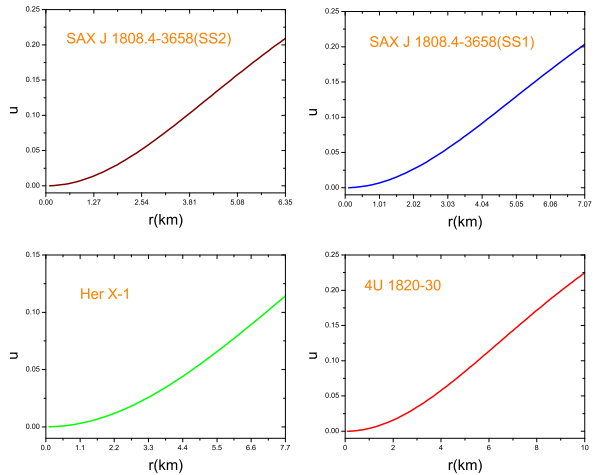}
\caption{Variation of u against radial parameter r of the star SAX J 1808.4-3658(SS2)(radius=6.35km), SAX J 1808.4-3658(SS1)(radius=7.07km), Her X-1(radius=7.7km) and 4U 1820-30(radius=10km).}
\label{fig:10}
\end{figure}

\begin{figure}[htbp]
    \centering
        \includegraphics[scale=.4]{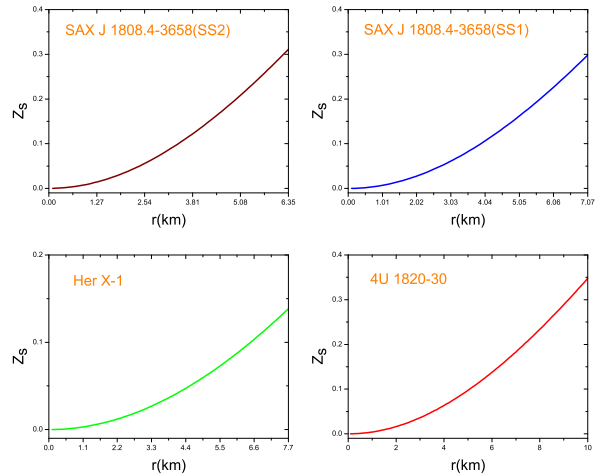}
        \caption{ Variation of the red-shift function($Z_{s}$) against radial parameter r of compact star SAX J 1808.4-3658(SS2)(radius=6.35km), SAX J 1808.4-3658(SS1)(radius=7.07km), Her X-1(radius=7.7km) and 4U 1820-30(radius=10km).}
    \label{fig:11}
\end{figure}

The surface redshift ($Z_s$) corresponding to the above
compactness ($u$) is obtained as
\begin{equation}
\label{eq36} 1+Z_s= \left[ 1-(2 u )\right]^{-\frac{1}{2}} ,
\end{equation}
where
\begin{equation}
\label{eq37} Z_s= \left[1-\frac{2ab^{2}}{R^{2}+2ab^{2}}\right]^{-\frac{1}{2}}-1
\end{equation}
Thus, the maximum surface redshift for the isotropic stars of different radii could be found very easily from FIG.~11 . We calculate the values of the different parameters in our model compact stars 4U 1820-30(radius=10km), Her X-1(radius=7.7km), SAX J 1808.4-3658(SS1)(radius=7.07km), SAX J 1808.4-3658(SS2)(radius=6.35km) from the above different equations (Table II). Whereas the standard values of different parameters of compact stars are given in Table III\citep{N. K. Glendenning1997,Gondek2000,Kalam2013b,Hossein2012}.

\begin{table*}
\begin{minipage}{140mm}
\caption{Values of the parameters evaluated for
different compact stars in our model.}\label{tbl}
\begin{tabular}{|llllllll|}
\hline
Compact Star & M($M_{\odot}$) & b(km) & $\rho_{0}$(gm/cc) & $\rho_{b}$(gm/cc) & $p_{0}$(dyne/cm$^2$) & $p_{b}$(dyne/cm$^2$) & $Z_{s}$ (max)  \\ \hline
SAX J 1808.4-3658(SS2) & 1.32806 & 6.35 & 28.722$\times10^{14}$ & 12.049$\times10^{14}$ & 4.125$\times10^{35}$ & 0.010$\times10^{35}$ & 0.311128  \\
SAX J 1808.4-3658(SS1) & 1.4379 & 7.07 & 22.094$\times10^{14}$ & 9.553$\times10^{14}$ & 3.027$\times10^{35}$ & 0.006$\times10^{35}$ & 0.298331  \\
Her X-1 & 0.879483 & 7.7 & 8.043$\times10^{14}$ & 5.261$\times10^{14}$ & 0.495$\times10^{35}$ & 0.003$\times10^{35}$ & 0.138451   \\
4U 1820-30 & 2.24719 & 10.0 & 13.148$\times10^{14}$ & 5.069$\times10^{14}$ & 2.13$\times10^{35}$ & 0.002$\times10^{35}$ & 0.34 \\
\hline
\end{tabular}
\end{minipage}
\end{table*}

\begin{table*}
\centering
\caption{Standard noted values of the parameters for
different compact stars.}\label{tbl}
\begin{tabular}{|llllll|}
\hline
Compact Star & $M$($M_{\odot}$) & $b$(km) & $\frac{M}{b}$ & $\rho_{0}$(gm/cc) & $p_{0}$(dyne/cm$^2$) \\ \hline
SAX J 1808.4-3658(SS2) & 1.323 & 6.35 & 0.308 & 28.348$\times10^{14}$ & 5.9830$\times10^{35}$ \\
SAX J 1808.4-3658(SS1) & 1.435 & 7.07 & 0.299 & 22.037$\times10^{14}$ & 4.16$\times10^{35}$ \\
Her X-1 & 0.88 & 7.7 & 0.168 & 8.2436$\times10^{14}$ & 0.8738$\times10^{35}$ \\
4U 1820-30 & 2.25 & 10.0 & 0.332 & 13.405$\times10^{14}$ & 1.9058$\times10^{35}$ \\
\hline
\end{tabular}
\end{table*}

\section{Results and Discussion}
In the present work we have investigated the physical nature of isotropic compact stars in the case of low -mass X-ray binarry 4U 1820-30 and X-ray pulsar SAX J 1808.4-3658(SS1), SAX J 1808.4-3658(SS2), Her X-1  considering the following : (a) The stars are isotropic in nature. (b) The space-time of the compact stars can be describe by M-W I metric.\\
We have obtained some interesting results which are as follows:\\ 
(i) At the centre, density and pressure of the steller interior is maximum and it decreses towards the surface (see Fig.1 - Fig.4), which are well behaved.\\
(ii) At the surface of the star, surface density and surface pressure have some finite value (see Table - I, II), but they vanishes out side of the star, which are physically acceptable.\\
(iii) It is to be noted that we have set Gaussian unit during solution of Einstein's equations. Incorporating SI unit to solve the equations, we calculate mass, central density, suface density and suface red-shift of strange stars, which are shown in the table - II.
Interestingly, we observe that the measured mass, central density, suface density, central pressure, surface pressure and suface red-shift of the taken strange stars are very much consistant with the reported data\citep{N. K. Glendenning1997,Gondek2000,Hossein2012,Kalam2013b}.\\
(iv) Our model is stable enough as it obey's the Herera's condition\citep{Herrera1992}.\\
(v) Any interior features of the stars can be obtain from mass-radius relation.\\
(vi) Adiabatic index, $\gamma > \frac{4}{3} $ is satisfied every where within the star (FIG.8).\\
(vii) Fig.6 clearly shows that, the static equlibrium configuration (TOV Equation) do exist due to the combined effect of gravitational ($F_{g}$) and hydrostatic ($F_{h}$) forces.\\
(viii) Our model obeys the dominant energy condition(DEC) i.e. $\rho_{0} > |p_{0}| $. It also satisfy all the limitations like $ \frac{mass}{radius}<{\frac{3}{8}}$  and surface red-shift$(Z_{s}\leq{1})$ given by D.E. Barraco and V.H. Hamity\citep{D.E. Barraco and V.H. Hamity2002} for isotropic star.\\
The entire analysis has been performed by a single metric for the different compact star candidates 4U 1820-30(radius=10km) in LMXBs and Her X-1(radius=7.7km), SAX J 1808.4-3658(SS1)(radius=7.07km), SAX J 1808.4-3658(SS2)(radius=6.35km) in X-ray pulsar. The analytical results confirms the validity of the propose model. Therefore, we suggest that this model will be widely applicable to the  compact stars in the low-mass X-ray binaries as well as in X-ray pulsars to investigate their physical properties.

\section*{Acknowledgments} MK, SMH gratefully acknowledge support
 from IUCAA, Pune, India under Visiting Associateship under which a part
  of this work was carried out.

\end{document}